# ENGINEERING AN ANTHROPOCENE CITIZENSHIP FRAMEWORK[1]

SHIMA BEIGI
*DEPARTMENT OF CIVIL ENGINEERING, UNIVERSITY OF BRISTOL, UNIVERSITY WALK, QUEEN'S BUILDING*
*BRISTOL, BS8 1TR, UK*

**ABSTRACT**
This article presents an Anthropocene citizen-cantered framework by incorporating the neuroscience of sustainability related stressors, the biology of collaboration in multi-agent ecosystems such as urban systems, and by emphasising on the importance of harnessing the collective intelligence of the crowd in addressing wicked challenges of sustainable development. The Anthropocene citizenship framework aims to transcend the cognitive model of global citizenship and sustainability to a dynamic, resilient and thriving mental model of collective cooperation.

**INTRODUCTION**
After Rio+20 Conference, the agenda of the Millennium Development Goals (MDGs) transformed into an outcome oriented vision and under the slogan of 'The Future We Want', the sustainable management of the links between people and the planet was tightly linked to improving communities resilience and urban systems' adaptability and also was shifted toward investing on building resilient infrastructure systems (Nations 2014).

Reaching these new challenges of resilient urbanisation, disaster management and infrastructure resilience are tightly interlinked. Therefore, one the one hand, the growing pace of urbanisation (Nations 2014) has to encourage engineering sustainable urban systems that have a global view on resilience and can deliver tangible outcomes. On the other hand, the interconnection between social, ecological and technical systems must also foster developing innovative solutions to climate change, poverty and hunger across the world with collaboration and collective action (Ogden, Heynen et al. 2013).

This article investigates ways to address the interconnection between sustainability goals through developing a novel resilience-oriented framework. The presented framework explores the impacts of uncertainty and complexity on communities' resilience.

To do so, in the first section a short review of resilience theory (Holling 1973) is presented and global view of the concept of resilience is introduced. In the second part of the paper, the author briefly reviews the phenomena of the Anthropocene (Kareiva, Lalasz et al. 2012) and major links that are affected by the consequences of this era is listed.

By identifying alternative pathways to address the challenges of the Anthropocene, a citizen-centric approach is suggested. The framework is focused on the behaviours of city planners' and citizens toward the challenges of living in complex urban systems as two fundamental pillars on the way to operationalise sustainability goals (Steffen, Persson et al. 2011).

---

## 1. A GLOBAL VIEW OF RESILIENCE

While initially resilience had been interpreted as a metaphor to address the complexity and interconnectivity of social and ecological systems (SES) (Holling 1996), it later become a fundamental shaper of systems' performance states under threats (Bruneau and Reinhorn 2004) and a determinant of systems' degree of adaptation, transformation and evolution (Carpenter, Walker et al. 2001, Walker, Holling et al. 2004). Across many disciplines, currently resilience is conceptualized as a transdisciplinary idea, a tool, a concept, and even an inspiration that encourages adopting a different approach to complexity and change (Masten 2001, Lance H. Gunderson and Holling 2002, Folke 2006, Rose 2007, White 2007, Cutter, Barnes et al. 2008). This growing rate of interest and thinking on the subject can shape a global resilience thinking (Walker and Salt 2006) framework e.g. see Figure 1. In this article, to create a vision for creating sustainable social-ecological and social-technical systems a global view of resilience is presented see Figure 1 (Beigi 2014). Figure 1 Beigi (2014) the following three points:

1) Social-ecological and social-technical systems are tightly interconnected (Levin 2005).
2) Without having a continuous and an uninterrupted access to flow of different types of resources and materials, these interconnected systems cannot be or achieve resilient. Therefore, maintaining a resilient access to the flow of these resources is essential for their smooth functioning of these systems and systemic management.
3) Social actors can change resilience at many levels and across many scales through their ability to think, self-organise and learn (e.g. the grey layer of cognitive map captures these capacities of social actors). Walker et al. (2004) call this meta-capacity as adaptability, defined as "the capacity of social actors to influence resilience".

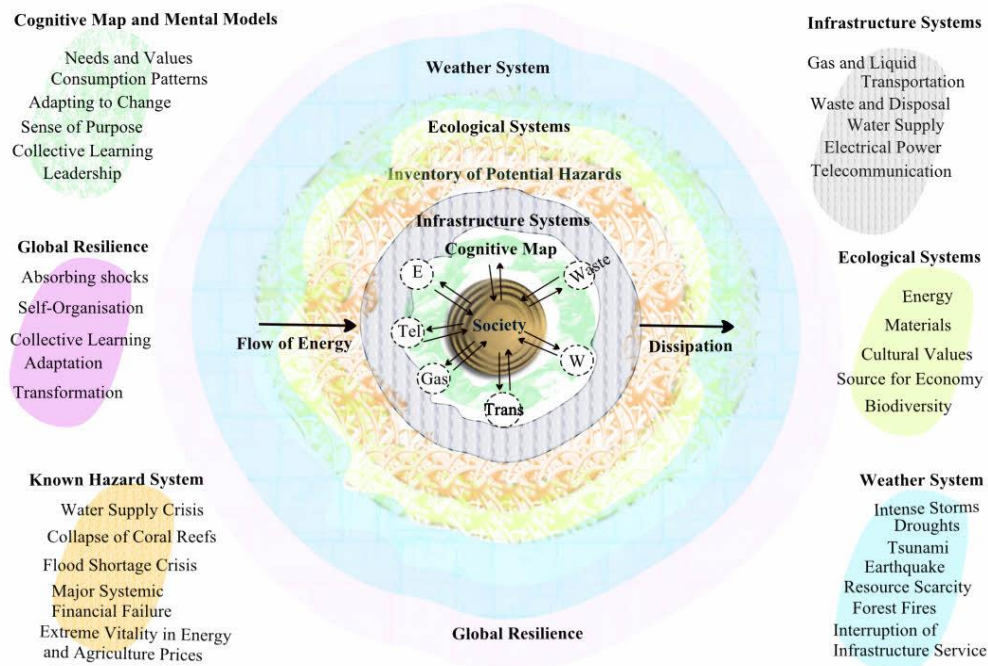

Figure 1. A global view of resilience. Adopted with permission from Beigi (2014).

## 2. THE ANTHROPOCENE CITIZENSHIP FRAMEWORK

The geological phenomenon of the Anthropocene is marked by the impacts of human behaviours (e.g. the process of urbanisation) on two major factors:

1. The stability and resilience of the Earth System (Rockstrom, Steffen et al. 2009, Steffen, Richardson et al. 2015).
2. The global resilience (e.g. see Figure 1) as one of the major concerns of the Anthropocene is on sustainability of ecosystem resources on which social-ecological and technical systems depend (Willows, Reynard et al. 2003).

The combination of these aspects influences the future of climate change mitigation programmes and the fulfilment of the millennium development goals (MDGS). In terms of MDGS and in particular the recent proposed goals of building resilient urban systems, it is of a particular importance to investigate how through building better urban systems the resilient of people and planet can be linked (Griggs, Stafford-Smith et al. 2013).

Since the majority of the research conducted on building and managing resilience tend to forget these links, because they are mainly focus on the negative impacts of this era on the planet (e.g. Path 1 , Figure 2), they often underestimate the opportunities that this unique time has to offer (e.g. Path 2 in Figure 2).

**WICKED PROBLEMS AND OUTCOME ORIENTED STRATEGIES**

To create an integrated framework, the Anthropocene citizenship framework is focused on building higher degree of adaptability to positively influence the global resilience (e.g. Figure. 3) and capturing the biological, psychological and physiological aspects of resilience and sustainability. The latter is developed through an investigation of the impacts of complexity and uncertainty on the processes of thinking, decisions making of social actors that result in adaptability within urban systems (e.g. see the cognitive map layer illustrated in Figure.1)

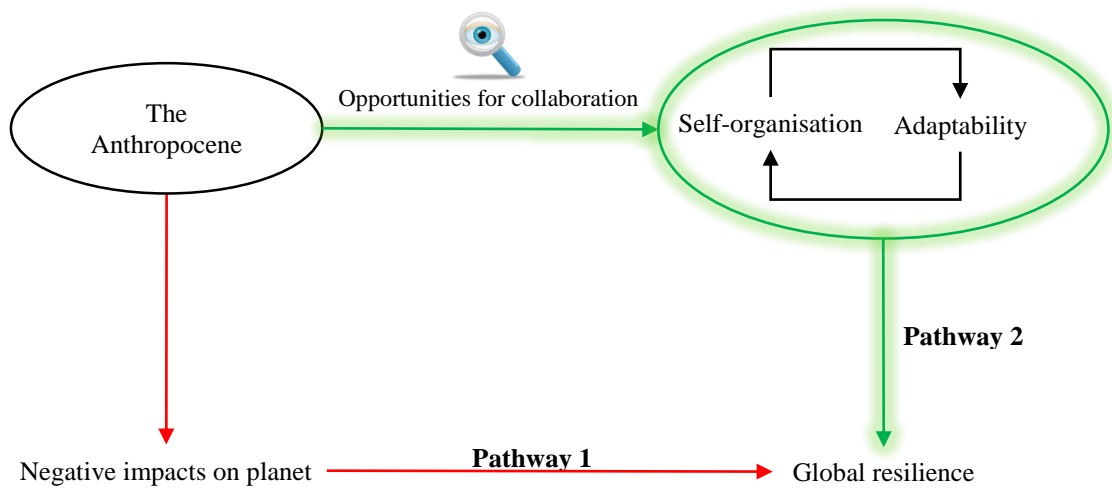

Figure 2. Author's depiction of the impacts and opportunities of the Anthropocene.

**HARNESSING THE INTELLIGENCE OF THE CROWD**

The events of the past decade such as tsunami of the Indian Ocean in 2004, the earthquakes in Haiti and Japan, and the hurricanes in the New Orleans and New York, are examples of complex problems that are often termed as wicked problems because they different parts and scales of systems.

The case of Japan's earthquake in the 11th of March 2011 is an instance of a wicked problem. The event resulted to the negative event of Fukushima Daiichi nuclear disaster. However, it also resulted in the emergence of community-based radioactivity monitoring activities which can be considered as a positive case where the interconnection and the subsequent shared risks between different parts of the social-ecological and technical system of Japan transformed the traditional process of disaster management in Japan (Aldrich 2012).

Another case that demonstrates the potential of the Anthropocene's for solving wicked problems such as sustainability and climate change in a collaborative way is the introduction of bus rapid transit (BRT) to the traditional network of Public transportation in Tehran, the capital of Iran following decades of struggling with the wicked problem of air pollution and congestion.

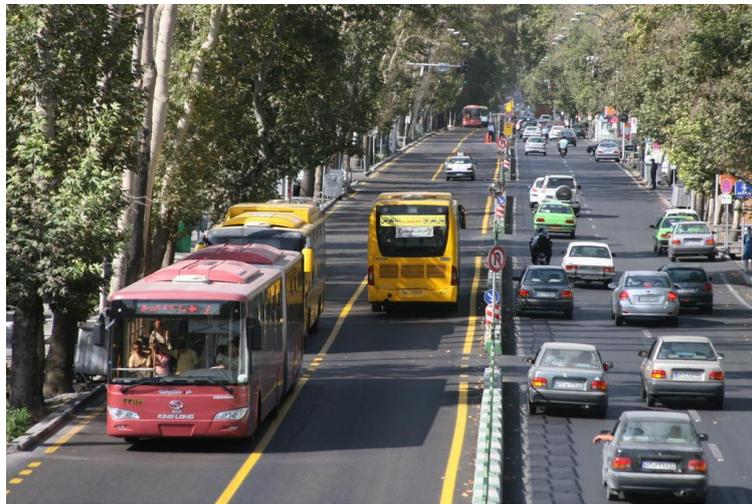

Figure 3. Bus rapid transit (BRT) lanes in Tehran, capital of Iran. Image courtesy of Twitter users.

The unique relevance of this case to the challenges of living in the Anthropocene is twofold. First, the proposed location for the development of the bus lanes. Valiasr, the proposed location, has cultural, social, and aesthetic dimensions where influence citizens' lives and behaviours deeply.

Second, because this unique street is the longest street in the Middle East, it connects critical regions of the city together. The challenge therefore was making sure different parts of the city of Tehran with their distinctive cultural facets collectively self-organise their commuting patterns in response to this new piece of infrastructure.

Due to history of air pollution, the urgent need for change led the authorities to investigate how by introducing a new piece of infrastructure that can delivers 'tangible and clear outcomes, citizens' cognitive model of transportation, culture, history and aesthetics can be linked.

As BRT *forced* people to change their commuting habits, it resulted in anger, frustration, traffic, and even political arguments among the citizens. However, with the introduction of technological tools such as Google Map, and platforms such as Swarm, the BRT is now a part of the citizen's culture of communication and transportation (Allen 2013). More importantly, BRT has become a platform for integrating women into the challenges of achieving sustainable development goals (e.g. See Allen (2013), section related to gender balance in BRT drivers).

The two cases of radioactivity monitoring in Japan and the challenge of addressing air pollution and congestion in Tehran display instances of opportunities for harnessing the intelligence of the crowd to provide solutions for wicked problems emerged from the interconnection between complex adaptive socio-ecological and socio-technical systems (Helbing 2013).

## BALANCING SELF-ORGANISATION AND ADAPTABILITY

Complex adaptive systems (CAS) such as social-ecological system, social-technical systems, and the human body have many interactive and adaptive parts. These systems exhibit properties such as self-organisation, feedback, nonlinearity, path-dependency, and emergence (Holland 1992, Levin 2005).

Self-organisation as an important part of the process of resilience is defined as: "The spontaneous emergence of an order without the intervention of an external force (Heylighen 1999)". The purpose of the framework is focused on building pathways that support proactive communities and moving citizens from low adaptability and low capacity for self-organisation toward the green quadrants.

Therefore, it formulates adaptability "the ability of social actors to influence resilience Walker et al. (2004)" and self-organisation simultaneously and investigate how the two can contribute to managing resilience and solving complex problems (Figure 5).

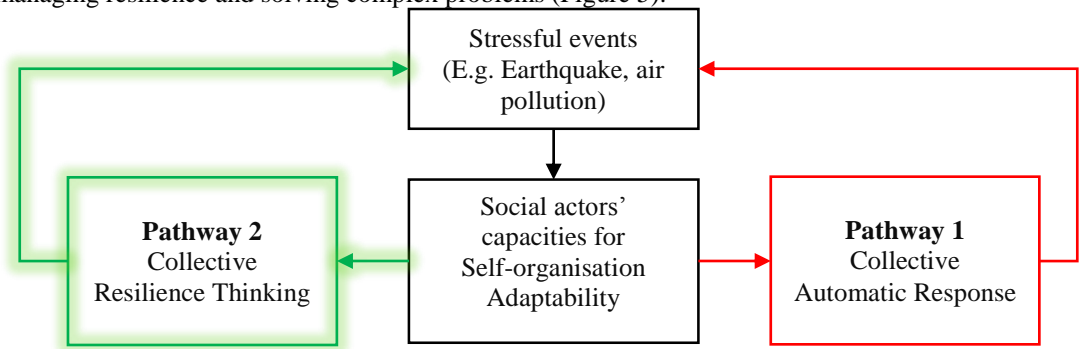

Figure 4. Author's depiction of the possible different response pathways to complex challenges.

**COOPERATIVE SELF-ORGANISATION REGIME**
The essence of the Anthropocene citizenship framework is enabling the whole of society to move to a regime of cooperative self-organisation. To actualise such regime, however, the processes through which social actors become aware of their impacts on their surround and on themselves have to be integrated in the developmental phase of sustainability oriented goals as lower adaptability capacities can hinder the path toward achieving the interconnected goals of sustainability.

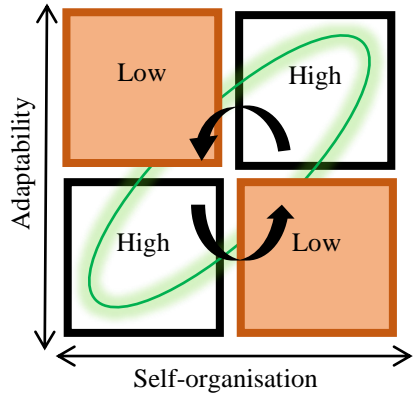

Figure 5. Author's depiction of the balance between self-organisation and adaptability.

Social systems whose adaptability is low 'have low capacity to influence their resilience' Walker et al. (2004), are not well-positioned to *autonomously* re-arrange the existing resources (i.e. self-organise), actively modify their behaviours, or adopt a new behaviour (e.g. transportation habits, recycling patterns, energy usage) (see Figure 5). Therefore, they are dependent on external factors (e.g. external fund, a catastrophic event, and top-down forces) which can lead to loss of trust across scales.

**CONCLUSION**
Complex factors such as loss of trust or conflicts between stakeholders and citizens are results of differences in the degrees of awareness of the scale of the challenges and of adaptability between parties which result in creation of resistance for change (e.g. the early phases of BRT in Tehran) and in extreme cases they can create poverty traps, where systems' remain in undesired conditions (Carter, Little et al. 2007).

To avoid such scenarios from developing, that because the interconnection between complex social-ecological and technical systems can happen rapidly and unexpected and even through surprise, the balance between adaptability and capacity for self-organisation needs explorations. Adaptability and self-organisation are initiated within social actors; the social actors in both cases must act on or respond to a task or condition.

It is essential to investigate how the complexity created by the challenges of the 21$^{st}$ century affects social actors' decisions and actions on their way toward resilience and building better prepared systems. Empowering social actors, engaging people in the process of decision making, and actively working with them through shaping the possible outcomes of adaptation strategies can create a low resistance pathway toward operationalising sustainable development goals.

**ACKNOWLEDGMEN**

The author would like to thank the listed funding bodies for supporting this research: Free University of Brussels (VUB), The Global Brain Institute (GBI), Professor Francis Heylighen and Professor Colin. A. Taylor, the University of Bristol (UOB), Engineering Physics Science Research Council (EPSRC), and the International Centre for Infrastructure Futures (ICIF).